\documentclass[12pt]{iopart}
\usepackage{cite}
\usepackage{algorithmic}
\usepackage{graphicx}
\usepackage{textcomp}
\usepackage{enumitem}
\usepackage{array}
\usepackage{diagbox}
\usepackage{url}

\begin{document}

\title{Robustness of link prediction under network attacks}

\author{Kun Wang, Lunbo Li and Cunlai Pu}

\address{School of Computer Science and Engineering, Nanjing University of Science and Technology, Nanjing 210094, China}
\ead{pucunlai@njust.edu.cn}
\vspace{10pt}
\begin{indented}
\item[]October 2018
\end{indented}

\begin{abstract}
While link prediction in networks has been a hot topic over the years, its robustness has not been well discussed in literature. In this paper, we study the robustness of some mainstream link prediction methods under various kinds of network attack strategies, including the random attack (RDA), centrality based attacks (CA), similarity based attacks (SA), and simulated annealing based attack (SAA).  Through the variation of precision, a typical evaluation index of link prediction, we find that for the SA and SAA, a small fraction of link removals can significantly reduce the performance of link prediction.  In general, the SAA has the highest attack efficiency, followed by the SA and then CA. Interestingly, the performance of some particular CA strategies, such as the betweenness based attacks (BA), are even worse than the RDA.  Furthermore, we discover that a link prediction method with high performance probably has lower attack robustness, and the vice versa.
\end{abstract}

%
\vspace{2pc}

\section{Introduction}\label{sec1}

The link prediction problem \cite{b42}, originated from the area of data mining, has achieved great progress with the booming of network science \cite{b51}. Its goal is to estimate the link possibility of two unconnected nodes based on the available network data and tools, such as machine learning \cite{b29,b30} and complex networks theory \cite{b52,b31}. Various link prediction methods haven been proposed \cite{b1,b53,b47,b48}, and they can be broadly divided into three categories: similarity-based methods (such as the common neighbors (CN) \cite{b42}, Adamic-Adar (AA) \cite{b8} and resource allocation (RA) \cite{b25} indices), maximum likelihood based methods (such as hierarchical structure model \cite{b9} and stochastic block model \cite{b10,b11}), and probabilistic models (such as probabilistic relational model \cite{b12} and probabilistic entity-relationship models \cite{b13}). The established link prediction methods can be widely used in online product recommendation \cite{b3}, bionetwork reconstruction \cite{b4}, community detection \cite{b5}, etc.

Though tremendous efficient prediction methods emerge, their robustness has not been seriously discussed in literature.  Unfortunately, real-world networks suffer from random failures and various targeted attacks. For instance, it has been shown that many scale-free networks, such as the Internet, are pretty vulnerable to degree-based targeted attacks \cite{b55,b35,b36}. A small initial attack can trigger the large-scale cascading failure \cite{b56,b57,b34}, which is one of the main security issues in power networks.  In addition, many novel attack strategies have been proposed, including edge attacks \cite{b39,b46}, path attacks \cite{b43,b44}, etc.  These random or intentional disturbances significantly affect the structure and dynamics of real-world networks \cite{b58}. In particular, the predictability of real-world networks keeps change as the network disturbance continues. Thus, it is very necessary to explore the robustness of link prediction under network attacks.

Recently, Zhang et al.\cite{b22} studied the robustness of several link prediction algorithms in noisy environments. Their results showed that while different prediction algorithms may have different attack robustness, in general they are robust to random disturbances. However, their work has the following limitations:

\begin{enumerate}[fullwidth,itemindent=1em]
\item Only the AUC is considered in the evaluation of prediction accuracy, while the precision is another mainstream index, which should be considered in the link prediction problem.
\item The perturbation methods used are relatively simple. Only random disturbances are considered. The robustness of link prediction under various targeted attacks needs more attention.
\end{enumerate}

In this paper, we study the robust of some typical link prediction methods under various network attacks.  Our main contributions are as follows:
\begin{enumerate}[fullwidth,itemindent=1em]
\item We investigated the link prediction robustness under the typical target attacks including the betweenness based attacks  (BA)  \cite{b28} and weight based attacks (WA) \cite{b41} and found that the WA has higher attack efficiency than the BA.
\item To further study the attack robustness, we proposed the similarity-based attack strategies by considering some typical similarity indices, such as the CN, AA and RA.  Furthermore, we proposed the simulated annealing based attack (SAA) strategy.
\item Through simulation, we found that the SAA has the largest attack efficiency, followed by the similarity-based strategies and then centrality-based strategies. Furthermore, we discovered that a link prediction method with high performance probably has lower robustness, and the vice versa.
\end{enumerate}

This paper is organized as follows. In the next section, we introduce the evaluation metrics as well as the  link prediction methods used in the paper. In Section \ref{sec3}, we provide the typical attack strategies and our proposed attack strategies. Section \ref{sec4} presents the experimental results and related analysis.  Finally, Section \ref{sec5} is our conclusion.

\section{Link prediction methods and their evaluation}\label{sec2}

We consider the structural similarity based  methods of link prediction, which are popular in the recent years \cite{b1}. In this type of methods, each unconnected node pair is given a similarity score according to a certain similarity index, and the node pairs of large similarity scores  have relatively large link probability.  Different from Ref. \cite{b22}, we use the precision, another typical metric,  to quantify the accuracy of link prediction.

\subsection{Link prediction methods}

We select five mainstream prediction methods, including the Common Neighbors (CN) \cite{b42}, Adamic-Adar (AA) \cite{b8}, Resource Allocation (RA) \cite{b25}, Local Path (LP) \cite{b26} and Katz \cite{b27}. In the following, the degree of  node $x$ is denoted by $k_x$, and  the set of common neighbors of nodes $x$ and $y$ is represented by $\Gamma(x)\cap\Gamma(y)$.

\begin{enumerate}[fullwidth,itemindent=1em]
\item CN: assuming the set of neighbor nodes of node $x$ to be  $\Gamma(x)$,   the CN index of  nodes $x$ and $y$ is defined as
\begin{equation}s^{CN}_{xy}=|\Gamma(x)\cap\Gamma(y)|.\end{equation}
\item AA: this index accounts for the contribution of common neighbors with a penalization term that is dependent on the logarithm of  common neighbors' degree,
\begin{equation}s^{AA}_{xy}=\sum_{z\in\Gamma(x)\cap\Gamma(y)}\frac{1}{\log{k_z}}.\end{equation}
\item RA: it is originated from  the resource allocation problem \cite{b40}. Suppose that node $x$  delivers resources to node $y$ (they are not directly connected) through their common neighbors. Each common neighbor  gets one unit of resources from $x$, and then evenly distributes the unit to all its neighbors.  The similarity between $x$ and $y$ can be defined as the amount of resources $y$ receives from $x$,
\begin{equation}s^{RA}_{xy}=\sum_{z\in\Gamma(x)\cap\Gamma(y)}\frac{1}{k_z}.\end{equation}
\item LP: this index  considers the contribution of paths of lengths 2 and 3,  which is defined as:
\begin{equation}\mathbf{S}^{LP}=\mathbf{A}^2+\alpha\cdot \mathbf{A}^3.\end{equation}
Where $\mathbf{A}$ is the adjacency matrix. The $(x,y)$ entry of $\mathbf{A}^i$ is the number of paths of length $i$ between nodes $x$ and $y$. $\alpha$ is a tunable parameter  controlling the contribution of paths with length 3.
\item Katz: it takes all the paths between two nodes into account, that is
\begin{equation}s^{Katz}_{xy}=\sum_{l=1}^\infty\alpha^l\cdot |paths_{x,y}^{<l>}|.\end{equation}
Where $\alpha$ is a tunable parameter and $|paths_{x,y}^{<l>}|$ represents the number of paths of length $l$ between nodes $x$ and $y$.
\end{enumerate}

\subsection{Evaluation metric}

We assume  an undirected and unweighted network $G(V,E)$, where $V$ is the set of nodes and $E$ is the set of links. In the link prediction problem, we  divide $E$ into a training set $E_T$ and a probe set $E_P$ to evaluate the prediction method. Usually, $E_P$ consists of 10\% (or 20\%) links randomly extracted from $E$, and $E_T$ includes the remaining 90\% (or 80\%) links. In $E_T$, each  unconnected node pair is given a similarity score by the similarity index, and then all the unconnected node pairs are ranked in decreasing order of similarity scores. The precision index for the evaluation of prediction accuracy is defined as
\begin{equation}precision=\frac{m}{L}.\end{equation}
Where $m$ is the number of desired node pairs, whose links are previously removed to the probe set,  among the top $L$ unconnected node pairs \cite{b1}.

\section{Network attack strategies}\label{sec3}

Real-world networks suffer from various kinds of failures or attacks. These disturbances should affect the  predictability of networks. To explore this, we consider the  representative attack strategies, including the random attack strategy (RDA) and centrality based attack strategies (CA) \cite{b58}. For the category of CA, we study the betweenness based attack (BA) \cite{b28} and the weight based attack (WA) \cite{b41}. Furthermore, we propose two additional types of attack strategies: similarity based attacks (SA) and  simulated annealing attack (SAA). Note that all the attack strategies mentioned here are for  link attacks, which further means removing the targeted links. All there strategies are described below in detail.
\begin{itemize}[fullwidth,itemindent=1em]
\item[(1)] RDA: the RDA is a very simple attack strategy, and usually works as a baseline of  other strategies. In the RDA, we randomly remove  links from the network.
\item[(2)] BA: in this strategy, we use betweenness centrality \cite{b28} to evaluate the importance of links. Then, we remove the links of the largest betweenness.  The betweenness of link $e$ is
\begin{equation}BC_e=\sum_{s,t\in V}\frac{\sigma(s,t|e)}{\sigma(s,t)}.\end{equation}
Where $\sigma(s,t)$ is the number of shortest paths between nodes $s$ and $t$, and $\sigma(s,t|e)$ is the number of those paths that also pass through link $e$.
\item[(3)] WA: in this strategy, we remove the links of the largest weight. Assume the end nodes of link $e$ are nodes $x$ and $y$,  the weight of link $e$ is given as \cite{b41}:
\begin{equation}W_e=k_x\cdot k_y.\end{equation}
Where $k_x$ and $k_y$ represent the degree of nodes $x$ and $y$, respectively.
\item[(4)] SA: for this category of attacks, the importance of a link is quantified as the similarity score of its two end nodes. Then, we remove the links of the largest importance. The CN, RA, and AA similarity indices are used here, and the corresponding attacks are denoted by SA-CN, SA-RA and SA-AA, respectively.   Note that the other similarity indices can also be taken into account.
\item[(5)] SAA: this strategy is based on the simulated annealing algorithm\cite{b23}. It globally searches the best set of links, by removing which the link prediction accuracy of the network decreases at the most.  The main procedures of the algorithm are as follows.
\begin{enumerate}[fullwidth,itemindent=1em]
\item we randomly select a set of links of a fixed number from the network as the initial solution, and set the   initial maximum temperature $t_{max}$ and minimum temperature $t_{min}$.
\item we randomly select a link from the  network (except the current solution) and another link from the current solution, and then exchange them to generate a new solution. Let $P_c$ ($P_n$) be the precision of the remaining network corresponding to the  current (new) solution.
\item If $P_n<P_c$, we replace the current solution with the new solution with probability 1, and otherwise  with probability $P=\exp{((P_c-P_n)/t_c)}$, where $t_c$ is the current temperature.
\item we update the temperature, $t_c=c\cdot t_c$, where $c$ is the cooling coefficient and has range in $(0,1)$. If $t_c<t_{min}$, the algorithm ends; otherwise, it returns to the second step.
\end{enumerate}
\end{itemize}
For the SAA, we remove the optimal set of links from the network.

\section{Result}\label{sec4}

We do experiments on various real-world network data downloaded from Refs.\cite{b49,b50},  including USAir (American Airlines Network), PB (Political Blog Network), C.elegans (Nematode Neural Network), Metabolic (Nematode Metabolism Network), Jazz (Jazz Musician Network) and Email (Email Communication Network). The statistics of these networks are shown in Table \ref{tab1}. Note that we ignore the link directions and remove self-loops in the network data. When calculating the precision, we do 100 times of training and test set divisions and get the average value. The results of RDA and SAA are the average of 10  independent runs.
\begin{table}
\caption{The statistics of six real-world networks, where $N$ is the number of nodes, $E$ is the number of links, and $C$ is the assortativity coefficient.}
\label{tab1}
\begin{indented}
\item[]\begin{tabular}{@{}ccccccc}
\br
 &
USAir&
PB&
C.elegans&
Metabolic&
Jazz&
Email\\
\mr
$N$&
332&
1,222&
297&
453&
198&
1,133\\
$E$&
2,126&
16,714&
2,148&
2,025&
2,742&
5,451\\
$C$&
-0.208&
-0.221&
-0.163&
-0.226&
0.020&
0.078\\
\br
\end{tabular}
\end{indented}
\end{table}

\subsection{Comparison of efficiency between different attack strategies}
\begin{figure*}\centering\includegraphics{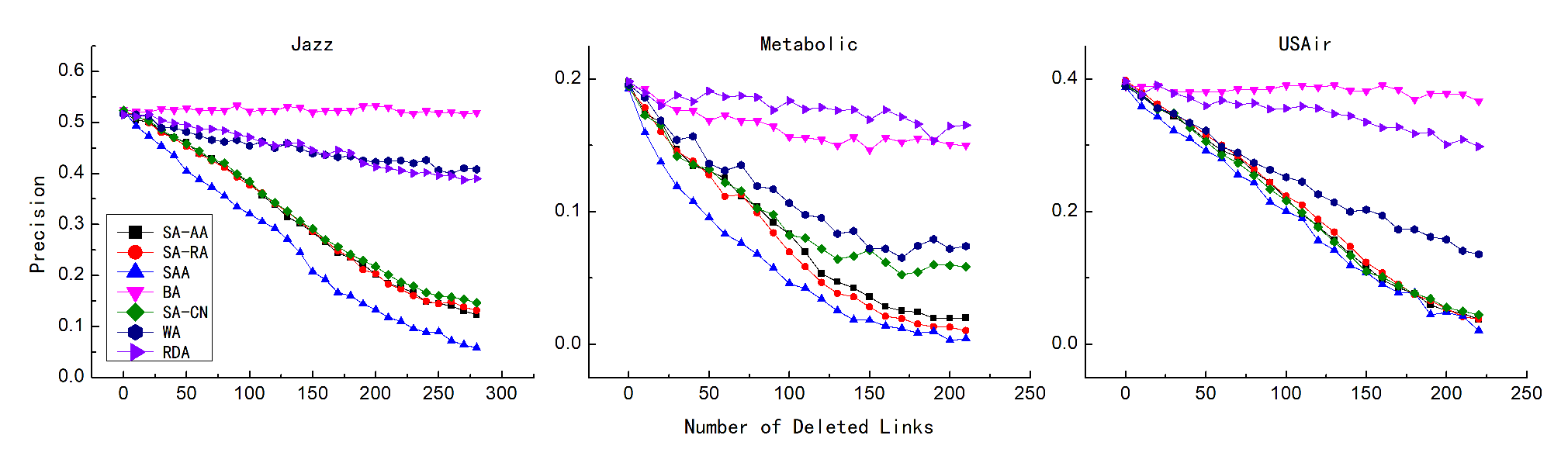}
\caption{Precision of the AA index under various attack strategies.}\label{fig1}
\end{figure*}

We first compare the  efficiency of the given attack strategies in terms of link prediction.  The AA index is used as the prediction method in the experiments.  The results  are shown in  Fig. \ref{fig1}. In each panel, the first data points  are corresponding to the case of no attacks, whereas the last data points are the results that 10\% of  links  are removed.

From this figure we can see that the SAA achieves the highest attack efficiency than the others, since it always attacks the optimal link set to suppress the precision. The SA is in the second place. The relative high efficiency of the SA lies in that removing the links, whose two end nodes are structurally similar, will greatly damage the structural similarity of the whole network, which further leads to the decrease of link predictability.  Note that the difference of efficiency between distinct SA strategies is not significant.

\begin{figure*}\centering\includegraphics[width=0.6\textwidth]{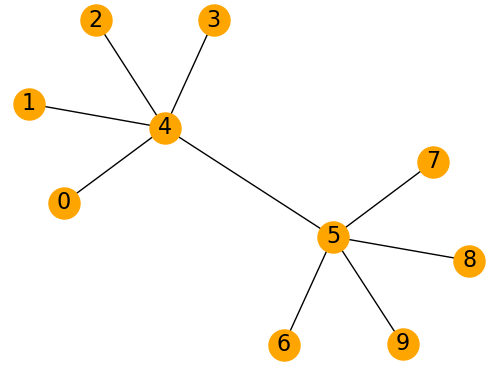}
\caption{An example network, in which link (4, 5) has the largest betweenness.\label{fig2}}
\end{figure*}

The CA follows the SA.  For different CA strategies, the efficiency is different. For instance, the WA is more efficient than the BA, which in some cases is even worse than the RDA. This is a little bit surprising, because the BA strategy is shown to be very efficient in other scenarios \cite{b39}. If a link has large betweenness, it is the intersection of many shortest paths, controlling the connectivity and communicability of the network. For example, link $(4,5)$ in Fig. \ref{fig2} has the largest betweenness and is critical to the network, since it connects two separate communities. However, nodes 4 and 5 have no common neighbor. Thus, they have a very small connection probability according to the rule of similarity based link prediction.  Removing this kind of links (outliers) does not affect the link predictability, and sometimes may even favor link prediction, since these links may be kind of noise to link prediction.

\begin{figure*}\centering\includegraphics{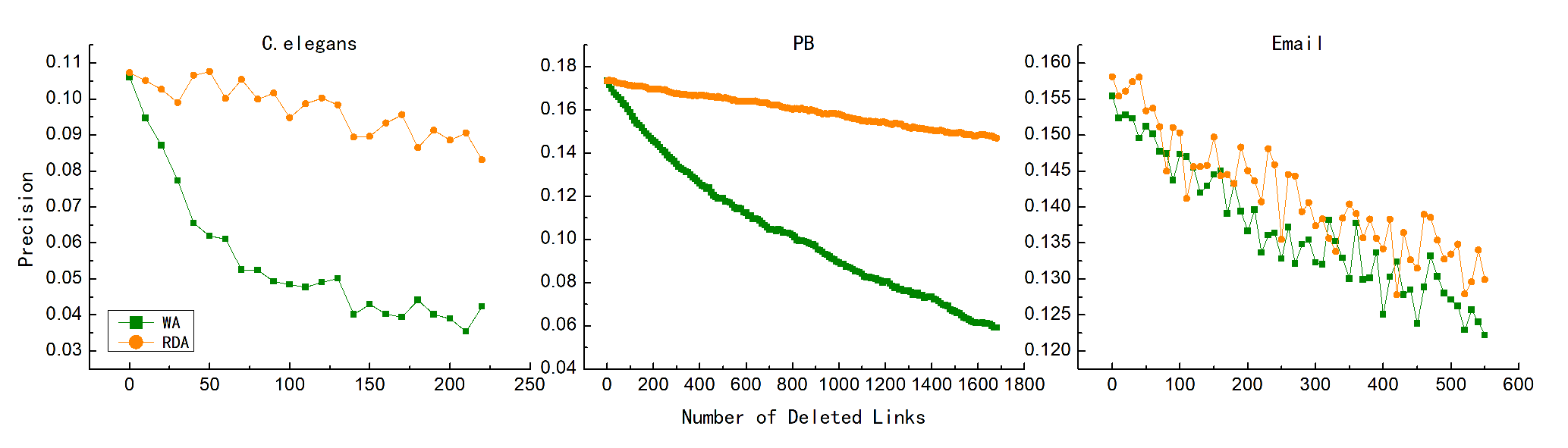}
\caption{Comparison of WA and RDA in different real-world networks.\label{fig4}}
\end{figure*}

For the WA, the attack efficiency of assortative networks are much larger than the disassortative networks, which is indicated in Fig. \ref{fig1} and Fig. \ref{fig4}. Note that the assortativity of the networks used in the experiments is given in Table \ref{tab1}, where  positive value means assortative network, and the vice versa.

To quantify the efficiency of an attack strategy, we further study the fraction of links needed to be removed in order to decrease the precision by half, which is named as attack cost and can be expressed as
\begin{equation}\Phi=\frac{E'}{E}.\end{equation}
Where $E'$ is the number of links that need to be removed to decrease the precision by half,  and $E$ is the total number of links in the network. based on the definition, we know that the smaller $\Phi$, the larger attack efficiency. We still use the AA index as the link prediction method.   $\Phi$ of three real-world networks under different attack strategies is given in  Table \ref{tab2}, in which ``-'' means $\Phi>10\%$.
\begin{table}
\caption{$\Phi$ (\%) of three real-world networks. If $\Phi>10\%$, we mark it as ``-''.\label{tab2}}
\begin{indented}
\item[]\begin{tabular}{@{}cccccccc}
\br
 &
BA&WA&SA-CN&SA-AA&SA-RA&SAA&RDA\\
\mr
Jazz&-&-&6.20&6.20&6.20&5.11&-\\
Metabolic&-&5.43&4.94&4.44&4.44&2.47&-\\
USAir&-&7.53&5.64&5.17&5.64&5.17&-\\
\br
\end{tabular}
\end{indented}
\end{table}
As can be seen from this table, for most of the attack strategies, removing only a small portion of  links can reduce the prediction accuracy by half, which indicates that the influence of  network attacks on link prediction cannot be ignored. In other words, real-world networks are  pretty vulnerable to network attacks in terms of link prediction. In addition, we can clearly see the difference of the efficiency of distinct attack strategies.

\subsection{Comparison of robustness between different prediction methods}
\begin{figure*}\centering\includegraphics{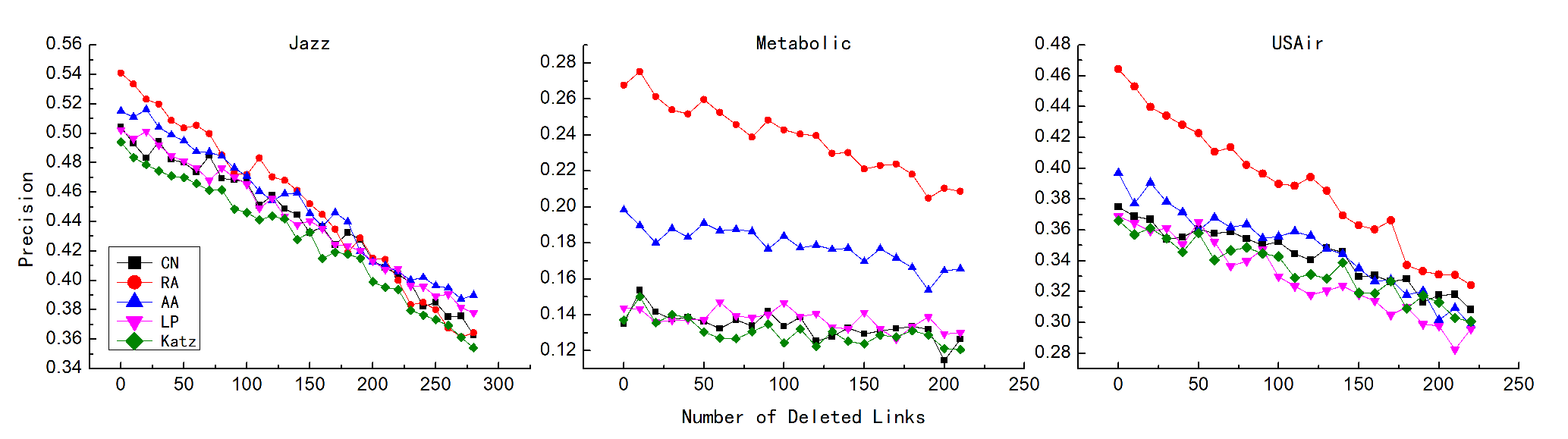}
\caption{Precision of different link prediction methods under the RDA attack strategy.\label{fig3}}
\end{figure*}

On the other hand, we compare the robustness of different link prediction methods. To be fair, we use the RDA as the attack strategy for all the prediction methods.  The experimental results are shown in Fig. \ref{fig3}. As in Fig. \ref{fig1}, the first data point of each curve corresponds to the case of no attacks, and the last one is the result when 10\% links are randomly removed. From Fig. \ref{fig3}, we observe that the precision of  all the link prediction methods decreases as the number of removed links increases. For the considered network data, the RA has the largest prediction accuracy, followed by the AA. The prediction accuracy of CN, LP and Katz is relatively low and with small difference.

In order to quantify the robustness of a  link prediction method against network attacks, we propose a new index of attack robustness as follows:
\begin{equation}\Omega=\frac{1}{L}\cdot\sum_{i=1}^L\frac{|precision(i\cdot \Delta m)-precision(0)|}{precision(0)}.\end{equation}
Where $\Delta m$ is a given constant, and $i$ is a variable number and has  range in $ [0,L]$.  $precision(*)$  is the result of precision when $*$ links are removed.  $\Omega$ measures the  average decrease rate of precision under network attacks, and small  $\Omega$ value means large attack robustness.

With the given definition of attack robustness index and the results of Fig. \ref{fig3},  we calculate  $\Omega$ of the five considered link prediction methods, the results of which are shown in  Table \ref{tab3}.
\begin{table}
\caption{$\Omega$ (\%) of link prediction methods under different real-world networks.\label{tab3}}
\begin{indented}
\item[]\begin{tabular}{@{}cccccc}
\br
 &CN&AA&RA&LP&Katz\\
\mr
Jazz&12.19&12.12&15.92&11.64&12.84\\
Metabolic&3.78&9.63&10.64&4.62&5.81\\
USAir&7.91&11.54&15.68&10.21&8.11\\
\br
\end{tabular}
\end{indented}
\end{table}
 We can see that although the RA has the highest prediction accuracy,  its $\Omega$ value is also the largest among all the prediction methods, which indicates that the RA has the smallest attack robustness.  In contrast, the prediction accuracy of CN is not very prominent. However,  its $\Omega$ value is smaller than the RA, which means that the CN is more robust than RA against network attacks.  These results call for a comprehensive consideration of the performance of a link prediction method in application.

\section{Conclusion}\label{sec5}
In summary, we study the robustness of link prediction methods under network attacks, a new dimension for the evaluation of  link prediction methods. Specifically, we consider the random attacks (RDA) and centrality based attacks (CA), i.e., the betweenness and weight based link attacks (BA and WA). Furthermore, we propose the similarity based attacks (SA), which attack the links based on their similarity scores, and the simulated annealing based attack (SAA).

Through the experiments, we observe that network attacks have a great impact on the link prediction accuracy, measured by the prediction index. Generally, attacking a small portion of links can result in a significant decrease of prediction accuracy, except for the random and betweenness based attacks. For all the attack strategies, the SAA has the highest attack efficiency, followed by the SA, and then CA and RDA. Note that in the category of centrality based attacks, the BA has low attack efficiency, even worse than the RDA.

For all the considered similarity indices, we obtain that the RA can achieve better prediction performance than the others, but it is very vulnerable to network attacks. On the contrary, the CN has relatively large robustness against attacks compared to the others, although its prediction performance may be just acceptable. These results indicate that the robustness of link prediction methods should also be considered, when they are used in real applications.

\section*{References}
\bibliographystyle{unsrt}  
\bibliography{refrences}

\begin{thebibliography}{10}

\bibitem{b42}
David Liben-Nowell and Jon Kleinberg.
\newblock The link-prediction problem for social networks.
\newblock {\em Journal of the American society for information science and
  technology}, 58(7):1019--1031, 2007.

\bibitem{b51}
Albert-L{\'a}szl{\'o} Barab{\'a}si et~al.
\newblock {\em Network science}.
\newblock Cambridge university press, 2016.

\bibitem{b29}
C.~Fu, M.~Zhao, L.~Fan, X.~Chen, J.~Chen, Z.~Wu, Y.~Xia, and Q.~Xuan.
\newblock Link weight prediction using supervised learning methods and its
  application to yelp layered network.
\newblock {\em IEEE Transactions on Knowledge and Data Engineering},
  30(8):1507--1518, Aug 2018.

\bibitem{b30}
Zheng Chen, Minmin Chen, Kilian~Q Weinberger, and Weixiong Zhang.
\newblock Marginalized denoising for link prediction and multi-label learning.
\newblock In {\em AAAI}, pages 1707--1713, 2015.

\bibitem{b52}
Aaron Clauset, Cristopher Moore, and Mark~EJ Newman.
\newblock Hierarchical structure and the prediction of missing links in
  networks.
\newblock {\em Nature}, 453(7191):98, 2008.

\bibitem{b31}
Linyuan L{\"u}, Liming Pan, Tao Zhou, Yi-Cheng Zhang, and H~Eugene Stanley.
\newblock Toward link predictability of complex networks.
\newblock {\em Proceedings of the National Academy of Sciences},
  112(8):2325--2330, 2015.

\bibitem{b1}
Linyuan Lü and Tao Zhou.
\newblock Link prediction in complex networks: A survey.
\newblock {\em Physica A: Statistical Mechanics and its Applications},
  390(6):1150 -- 1170, 2011.

\bibitem{b53}
V{\'\i}ctor Mart{\'\i}nez, Fernando Berzal, and Juan-Carlos Cubero.
\newblock A survey of link prediction in complex networks.
\newblock {\em ACM Computing Surveys (CSUR)}, 49(4):69, 2017.

\bibitem{b47}
Fei Tan, Yongxiang Xia, and Boyao Zhu.
\newblock Link prediction in complex networks: A mutual information
  perspective.
\newblock {\em PLOS ONE}, 9(9):1--8, 09 2014.

\bibitem{b48}
Boyao Zhu and Yongxiang Xia.
\newblock An information-theoretic model for link prediction in complex
  networks.
\newblock {\em Scientific Reports}, 5:13707, 2015.

\bibitem{b8}
Lada~A Adamic and Eytan Adar.
\newblock Friends and neighbors on the web.
\newblock {\em Social Networks}, 25(3):211--230, 2003.

\bibitem{b25}
Tao Zhou, Linyuan Lü, and Yi~Cheng Zhang.
\newblock Predicting missing links via local information.
\newblock {\em European Physical Journal B}, 71(4):623--630, 2009.

\bibitem{b9}
Marta Sales-Pardo, Roger Guimera, Andr{\'e}~A Moreira, and Lu{\'\i}s A~Nunes
  Amaral.
\newblock Extracting the hierarchical organization of complex systems.
\newblock {\em Proceedings of the National Academy of Sciences},
  104(39):15224--15229, 2007.

\bibitem{b10}
Paul~W. Holland, Kathryn~Blackmond Laskey, and Samuel Leinhardt.
\newblock Stochastic blockmodels: First steps.
\newblock {\em Social Networks}, 5(2):109--137, 1983.

\bibitem{b11}
Edoardo~M Airoldi, David~M Blei, Stephen~E Fienberg, and Eric~P Xing.
\newblock Mixed membership stochastic blockmodels.
\newblock {\em Journal of Machine Learning Research}, 9(Sep):1981--2014, 2008.

\bibitem{b12}
Jennifer Neville.
\newblock {\em Statistical models and analysis techniques for learning in
  relational data}.
\newblock University of Massachusetts Amherst, 2006.

\bibitem{b13}
David Heckerman, Chris Meek, and Daphne Koller.
\newblock Probabilistic entity-relationship models, prms, and plate models.
\newblock pages 55--60, 2007.

\bibitem{b3}
Greg Linden, Brent Smith, and Jeremy York.
\newblock Amazon.com recommendations: Item-to-item collaborative filtering.
\newblock {\em IEEE Internet Computing}, 7(1):76--80, 2003.

\bibitem{b4}
Markus~J Herrg{\aa}rd, Neil Swainston, Paul Dobson, Warwick~B Dunn, K~Yalcin
  Arga, Mikko Arvas, Nils Bl{\"u}thgen, Simon Borger, Roeland Costenoble,
  Matthias Heinemann, et~al.
\newblock A consensus yeast metabolic network reconstruction obtained from a
  community approach to systems biology.
\newblock {\em Nature biotechnology}, 26(10):1155, 2008.

\bibitem{b5}
Filippo Radicchi, Claudio Castellano, Federico Cecconi, Vittorio Loreto, and
  Domenico Parisi.
\newblock Defining and identifying communities in networks.
\newblock {\em Proceedings of the National Academy of Sciences of the United
  States of America}, 101(9):2658--2663, 2004.

\bibitem{b55}
R{\'e}ka Albert, Hawoong Jeong, and Albert-L{\'a}szl{\'o} Barab{\'a}si.
\newblock Error and attack tolerance of complex networks.
\newblock {\em nature}, 406(6794):378, 2000.

\bibitem{b35}
Paolo Crucitti, Vito Latora, Massimo Marchiori, and Andrea Rapisarda.
\newblock Efficiency of scale-free networks: error and attack tolerance.
\newblock {\em Physica A Statistical Mechanics \& Its Applications},
  320(C):622--642, 2002.

\bibitem{b36}
Lazaros~K. Gallos, Reuven Cohen, Panos Argyrakis, Armin Bunde, and Shlomo
  Havlin.
\newblock Stability and topology of scale-free networks under attack and
  defense strategies.
\newblock {\em Phys. Rev. Lett.}, 94:188701, May 2005.

\bibitem{b56}
Yang Yang, Takashi Nishikawa, and Adilson~E Motter.
\newblock Small vulnerable sets determine large network cascades in power
  grids.
\newblock {\em Science}, 358(6365):eaan3184, 2017.

\bibitem{b57}
Benjamin Sch{\"a}fer, Dirk Witthaut, Marc Timme, and Vito Latora.
\newblock Dynamically induced cascading failures in power grids.
\newblock {\em Nature communications}, 9(1):1975, 2018.

\bibitem{b34}
Ye~Cai, Yijia Cao, Yong Li, Tao Huang, and Bin Zhou.
\newblock Cascading failure analysis considering interaction between power
  grids and communication networks.
\newblock {\em IEEE Transactions on Smart Grid}, 7(1):530--538, 2016.

\bibitem{b39}
Petter Holme, Beom~Jun Kim, Chang~No Yoon, and Seung~Kee Han.
\newblock Attack vulnerability of complex networks.
\newblock {\em Phys. Rev. E}, 65:056109, May 2002.

\bibitem{b46}
Igor Mishkovski, Mario Biey, and Ljupco Kocarev.
\newblock Vulnerability of complex networks.
\newblock {\em Communications in Nonlinear Science and Numerical Simulation},
  16(1):341 -- 349, 2011.

\bibitem{b43}
Cun-Lai Pu and Wei Cui.
\newblock Vulnerability of complex networks under path-based attacks.
\newblock {\em Physica A: Statistical Mechanics and its Applications}, 419:622
  -- 629, 2015.

\bibitem{b44}
Cunlai Pu, Siyuan Li, Andrew Michaelson, and Jian Yang.
\newblock Iterative path attacks on networks.
\newblock {\em Physics Letters A}, 379(28):1633 -- 1638, 2015.

\bibitem{b58}
Reuven Cohen and Shlomo Havlin.
\newblock {\em Complex networks: structure, robustness and function}.
\newblock Cambridge university press, 2010.

\bibitem{b22}
Peng Zhang, Xiang Wang, Futian Wang, An~Zeng, and Jinghua Xiao.
\newblock Measuring the robustness of link prediction algorithms under noisy
  environment.
\newblock {\em Scientific reports}, 6:18881, 2016.

\bibitem{b28}
Ulrik Brandes.
\newblock On variants of shortest-path betweenness centrality and their generic
  computation.
\newblock {\em Social Networks}, 30(2):136 -- 145, 2008.

\bibitem{b41}
Rui Yang, Wen-Xu Wang, Ying-Cheng Lai, and Guanrong Chen.
\newblock Optimal weighting scheme for suppressing cascades and traffic
  congestion in complex networks.
\newblock {\em Phys. Rev. E}, 79:026112, Feb 2009.

\bibitem{b26}
Linyuan L{\"u}, Ci-Hang Jin, and Tao Zhou.
\newblock Similarity index based on local paths for link prediction of complex
  networks.
\newblock {\em Physical Review E}, 80(4):046122, 2009.

\bibitem{b27}
Leo Katz.
\newblock A new status index derived from sociometric analysis.
\newblock {\em Psychometrika}, 18(1):39--43, 1953.

\bibitem{b40}
Qing Ou, Ying-Di Jin, Tao Zhou, Bing-Hong Wang, and Bao-Qun Yin.
\newblock Power-law strength-degree correlation from resource-allocation
  dynamics on weighted networks.
\newblock {\em Phys. Rev. E}, 75:021102, Feb 2007.

\bibitem{b23}
S.~Kirkpatrick, C.~D. Gelatt, and M.~P. Vecchi.
\newblock Optimization by simulated annealing.
\newblock {\em Science}, 220(4598):671--680, 1983.

\bibitem{b49}
Link prediction group.
\newblock \url{http://www.linkprediction.org/}.

\bibitem{b50}
Konect.
\newblock \url{http://konect.uni-koblenz.de/networks/arenas-email}.

\end{thebibliography}

\end{document}